\documentstyle[12pt,psfig]{article} 

\newcommand{\be}{\nopagebreak\begin{equation}} 
\newcommand{\ee}{\end{equation}} 
\newcommand{\ba}{\begin{array}} 
\newcommand{\ea}{\end{array}} 
\newcommand{\bp}{\begin{picture}} 
\newcommand{\ep}{\end{picture}}

\newcommand{\rf}[1]{Ref.~\cite{#1}} 
\newcommand{\eq}[1]{Eq.~(\ref{#1})} 
\newcommand{\bi}[6]{\bibitem{#1}{#2, }{\sl #3 }{\bf #4}{ (#5)}{ #6}} 
\newcommand{\wt}[1]{\widetilde{#1}}

\newcommand{\eg}{{\it e.g.\ }} 
\renewcommand{\d}{\partial} 
\newcommand{\tr}{{\rm tr}\;}

\newcommand{\scs}{\scriptscriptstyle} 
\newcommand{\nnl}{\noindent} 
 
\newcommand{\mv}[1]{\langle #1 \rangle}

\newcommand{\newsection}{    % Numeration of eqs. is automatic 
\setcounter{equation}{0} 
\section}

\begin{document} 

\begin{titlepage} 

\begin{flushright} 

EPHOU-00-001\\
%Preprint ITEP 13/98\\
NBI-HE-00-09\\
%MPS-RR-1998-xx\\
TIT/HEP-437\\
Jan. 2000\\
\end{flushright} 

\vspace*{5pc} 

\begin{center} 

{\LARGE \bf Recursive sampling  simulations of \\
3D gravity coupled to scalar fermions}
\end{center} 

\vspace{1.5pc} 

\begin{center} 

 {\large J. Ambj\o rn}\\ 

%\vspace{0.5pc} 

{\em Niels Bohr Institute\\ 
Blegdamsvej 17, Copenhagen \O, Denmark}\\ambjorn@nbi.dk\\ 

\vspace{1pc} 

 {\large D.V. Boulatov}\\ 
%\vspace{0.5pc} 
{\em Institute for Theoretical and Experimental Physics (ITEP)\\ 
B.Cheremushkinskaya 25, Moscow, Russia}\\boulatov@heron.itep.ru\\ 

\vspace{1pc} 

 {\large N. Kawamoto}\\ 
%\vspace{0.5pc} 
{\em Physics Department, Hokkaido University\\ 
Sapporo, Japan} 
\\kawamoto@particle.sci.hokudai.ac.jp\\ 

\vspace{1pc} 

 {\large Y. Watabiki}\\ 
%\vspace{0.5pc} 
{\em Tokyo Institute of Technology\\ 
Meguro, Tokyo, Japan}\\ watabiki@th.phys.titech.ac.jp

\end{center} 

\vspace{1pc} 

\begin{center} 

{\large\bf Abstract} 

\end{center} 

We study numerically the phase structure of a model of 3D gravity 
interacting with scalar fermions.  
We measure the 3D counterpart of the ``string" susceptibility exponent 
as a function of the inverse Newton coupling $\alpha$. 
We show that there are two phases separated by a critical point 
around $\alpha_c \simeq 2$.
The numerical results support the hypothesis that 
the phase structures of 3D and 2D simplicial gravity 
are qualitatively similar, 
the inverse Newton coupling in 3D  playing the role 
of the central charge of matter in 2D.

\vfill

\end{titlepage}

\newsection{Introduction} 
The success of the so-called dynamical triangulations or 
matrix models as a theory of 2D quantum gravity and  
non-critical strings has brought about a hope, that analogous discrete  
approaches might be useful in higher dimensions as well. The most  
natural way to introduce discretized quantum gravity is to consider  
simplicial complexes instead of continuous manifolds. Then, a path  
integral over metrics can be simply defined as a sum over all 
complexes having some fixed properties. 
For example, it is natural to restrict their topology.  
In the present paper, we consider only the 3-dimensional case, where,  
by definition, a simplicial complex is a collection of tetrahedra glued  
along their faces in such a way that any two of them can have at most one  
triangle in common. In addition we will restrict the topology to be 
that of the three-sphere.

To introduce metric properties, one can assume that complexes represent 
piece-wise linear manifolds constructed by gluing together equilateral 
tetrahedra \cite{Regge}. 
The volume is then proportional to the number of them.  
In the continuum limit, as this 
number grows, the edge length, $a$, simultaneously tends to zero: $a\to0$. In 
three dimensions, there is a one-to-one correspondence between classes of 
piece-wise linear and smooth manifolds. In other words, any homeomorphism can 
equally well be approximated by either a piece-wise linear or a smooth map. 
This property makes the model self-consistent. 

Within a piece-wise linear approximation, 
the curvature is singular: the space is flat everywhere except at the edges. 
Therefore, one should consider integrated quantities, 
\eg the mean scalar curvature 
\be \int\! d^3x\, \sqrt g R \ = \ a \Big(2\pi N_1 -6N_3\arccos \frac13\Big) \ee 
where $a$ is the lattice spacing and $N_0,\ N_1,\ N_2,\ N_3$ are the numbers 
of vertices, edges, triangles and tetrahedra, respectively. 

For manifolds in 3 dimensions, the Euler character vanishes $\chi = 
N_0-N_1+N_2-N_3=0$.  Together with the other constraint $N_2=2N_3$, it means 
that only 2 of the $N_i$'s are independent, $N_1=N_0+N_3$, and a natural 
lattice action depends on 2 dimensionless parameters. Now, we are in a 
position to define the simplicial-gravity partition function 
\cite{ADJ,3dsim,3dsim2,B}: 
\be 
{\bf Z}(\alpha,\mu) \ = \ \sum_{\{C\}} e^{\alpha N_0 - \mu N_3} 
\ = \  \sum_{N_3} Z_{N_3}(\alpha) e^{-\mu N_3} 
\label{Z} 
\ee
where $\sum_{\{C\}}$ is a sum over some class of complexes 
(\eg of a fixed topology, as mentioned above). 
$\alpha = 1/G$ is the inverse Newton constant 
and $\mu$ is the cosmological constant. 

It can be shown that if $N_3$ is fixed, $N_0$ is restricted from above by 
$N_0<\frac13N_3+const$. Therefore, it is natural, keeping $\alpha$ fixed, 
to let $\mu$ approach its critical value, $\mu_c$, at which the sum over $N_3$ 
in \eq{Z} becomes divergent. In the vicinity of $\mu_c$ one could hope 
to find critical behavior corresponding to a continuum limit of the model.
In general the value of $\mu_c$ will be a function, $\mu_c(\alpha)$, 
of the other parameter $\alpha$. While $\mu \to \mu_c(\alpha)$ in general 
will correspond to an infinite volume limit ($N_3 \to \infty$), 
it might require a fine-tuning of $\alpha$ to  find a limit
which has interesting {\em continuum} physics. From this point 
of view the parameter $\alpha$ might play a role similar to the one 
played by the (inverse) temperature $\beta$ in the Ising model. While we 
can take the infinite volume limit of the Ising model for any $\beta$, 
it is only for one specific value of $\beta$ (corresponding to the 
critical temperature) that we can relate the Ising model to 
a continuum $c=1/2$ conformal field theory.

For the model to be physically consistent, the free energy per unit volume, 
$\frac1{N_3}\log Z_{N_3}(\alpha)$, has to be finite. It implies that the 
series over $N_3$ in \eq{Z} has to have a finite radius of convergence: 
$0<\mu_c<+\infty$. If this holds, statistical models defined on such an 
ensemble of fluctuating lattices possess reasonable thermodynamical 
behavior. Thus one can hope to describe matter coupled to quantum 
gravity. This program has been successfully carried out in two dimensions 
with help of the matrix model techniques \cite{KKM,O(n)}. In the 3D 
case, analytic tools are sparse and numerical simulations have so far 
been the main source of information. In the case of pure 3D 
gravity in references \cite{3dsim,3dsim2}, and  in the case of matter
fields coupled to 3D gravity in references \cite{matter}. 

Many questions, which are simple in two dimensions, become complicated
 in three. For example, 2-manifolds are classified according to 
values of the Euler character, and therefore topology, homotopy and 
homology are actually equivalent descriptions. In three dimensions, 
sets of manifolds 
characterized according to these 3 criteria are essentially different. 
From this point of view it is natural to ask how large a class 
of complexes $\{C\}$ one can include in \eq{Z}, in order that the 
model still has acceptable properties, like being well defined and 
having a finite free energy per unit volume. 
It has been argued \cite{Bo} that it is sufficient to fix only the first 
homology group $H_1(C,Z)$ in order to have a finite value of $\mu_c$. It 
can be reformulated as the following statement: ``The number of simplicial 
manifolds constructed of the given number of tetrahedra, $N_3$, and having 
a fixed homology type ({\em viz.,} the first Betti number and the torsion 
coefficients) grows at most exponentially as a function of $N_3$''.  

Let us further note that the existence of such an exponential bound  
will remain valid if one attribute positive weights which grow
at most exponentially with $N_3$, to every complex. 
The other way around, if one manages to represent a sum 
over complexes as a weighted  
sum over another class of objects, the weights again exponentially bounded,
then it is sufficient to prove that the number of objects 
in the new class is exponentially bounded. 
In the class of regular triangulations there is firm 
numerical evidence for an exponential bound \cite{numbounds}. Further,
analytical arguments now exist, which strongly favor an exponential bound
\cite{anabound}.

The simplest example of a weighted sum of complexes is given by the 
canonical partition function $Z_{N_3}(\alpha)$ introduced in 
\eq{Z}. Another example of physical interest is the partition function in 
the presence of free matter fields. To introduce it let us consider the 
set of polyhedral complexes dual to simplicial ones. Their 1-skeletons are some 
4-valent graphs, $G^{(4)}$, whose vertices correspond to tetrahedra, and 
links are dual to triangles. Such a graph can be defined by the 
adjacency matrix 
\be  
G^{(4)}_{ij}=\left\{\ba{ll}  
1 & \mbox{if vertices $i$ and $j$ are connected by a link}\\  
0 & \mbox{otherwise}  
\ea \right.   
\ee 

Of course the set of 4-valent graphs is not identical to the totality of 
simplicial complexes. Similarly, in two dimensions, the set of ordinary 
$\phi^3$ Feynmann diagrams is different from that of triangulations -- 
one has to introduce the notion of fat graphs to establish the 
equivalence.   

If one attaches a $n$-component vector $x^{\mu}_i\  
(\mu=1,\ldots,n)$ to the $i$-th vertex, the corresponding gaussian integral  
can be performed explicitly for every given graph $G^{(4)}$: 
\be 
Z_{\rm matter}(G^{(4)}) \ = \
\int \prod_{i=1}^{N_3-1}\prod_{\mu=1}^n d x^{\mu}_i \exp\Big[ -\frac12  
\sum_{i,j=1}^{N_3-1}G^{(4)}_{ij}(x^{\nu}_i-x^{\nu}_j)^2\Big] 
\ = \ \Big(\det\mbox{}' L^{(4)}\Big)^{-\frac{n}2} \, ,
\label{gint} 
\ee 
\nnl where the discrete Laplacian is given by 
\be 
L^{(4)}_{ij} \ = \ 4 \delta_{ij}-G^{(4)}_{ij} \, .
\ee
There is the nice combinatorial representation of the determinant  
given by Kirchoff's theorem \cite{Biggs}: 
\be 
\det\mbox{}' L  \ = \ |T(G)| \, ,
\ee 
where $T(G)$ is the set of all possible spanning trees embedded into 
a graph $G$ or, equivalently, the number of connected trees which can be 
obtained from the graph by cutting its links.  To 
remove the zero mode in \eq{gint}, we fixed the field at the $N_3$'th 
vertex. Kirchoff's theorem states that $\det\mbox{}' L$ 
does not depend on the choice of 
the vertex. Therefore, $|T(G)|$ is the number of rooted trees. 

The number of spanning trees of a $k$-valent graph $G^{(k)}_n$ with $n$ 
vertices can be estimated from above as \cite{Biggs} 
\be 
|T(G^{(k)}_n)|  \ \leq \ \frac1n\Bigg(\frac{nk}{n-1}\Bigg)^{n-1} \, .
\ee 

The case we are particularly interested in is the 2-component Grassmann 
field (or equivalently, $n=-2$ in eq.\ (\ref{gint})), 
where we obtain the gravity+matter partition function in the form 
\be 
{\bf Z}^{(1)}(\alpha,\mu)
\ = \ \sum_{\{C\}}\det\mbox{}' L\;  
e^{\alpha N_0 - \mu N_3} 
\ = \ \sum_{N_3} Z^{(1)}_{N_3}(\alpha)e^{-\mu N_3} \, .
\label{Z1a} 
\ee 
The weights, $ \det\mbox{}' L$, are positive integers, hence, 
\be 
Z_{N_3}(\alpha) \ < \ Z^{(1)}_{N_3}(\alpha) \, .
\ee 

In two dimensions, this type of matter has the central charge $c=-2$ and the  
corresponding matrix model has been solved explicitly by purely 
combinatorial means \cite{KKM}.  
We can repeat the same trick in 3 dimensions: 
\be  
{\bf Z}^{(1)}(\alpha,\mu)
\ = \ \sum_{\{C\}}\det\mbox{}' L\; e^{\alpha N_0-\mu N_3} 
\ = \ \sum_{\{C\}}\sum_{\{\wt{T}(C)\}}e^{\alpha N_0-\mu N_3}
\ = \ \sum_{\{\wt{T}\}} \sum_{\{C(\wt{T})\}}e^{\alpha N_0-\mu N_3} .  
\label{Z1} 
\ee  
\nnl Here we have a sum over complexes, $\sum_{\{C\}}$,  
weighted with the number of spanning trees in 
the dual 1-skeleton of each complex, $\sum_{\{\wt{T}(C)\}}$.  
We can obtain exactly the 
same class of configurations by taking the sum over all possible trees, 
$\sum_{\{\wt{T}\}}$, weighted with the number of complexes,  
$\sum_{\{C(\wt{T})\}}$, 
which can be reconstructed from a given tree $\wt{T}$ by ``restoring'' 
links which we imagine have been cut in order to produce the tree
from the original graph. 

In three dimensions, the trees appearing this way correspond to spherical 
simplicial balls obtained starting with a single tetrahedron 
and subsequently gluing other tetrahedra to the faces of the boundary. 
If we denote by $n_0,\ n_1$ and $n_2$ the numbers of vertices, edges and  
triangles on the boundary of a ball, we find that  
\be 
n_0 = N_3+3 \, , \hspace{2pc} 
n_2 = 2(N_3+1) \, .
\ee 

It is convenient to weight every complex with the number of spanning 
trees $T$. This is an exponential reweighting of the kind discussed
above. It can be done by introducing another 
system of free matter fields attached to vertices of triangulations. Then, 
analogously to \eq{Z1}, we define the modified partition function 
\be  
{\bf Z}^{(2)}(\alpha,\mu)
\ = \ \sum_{\{C\}} \det\mbox{}'L \; 
      \det\mbox{}'\wt{L} \; e^{\alpha N_0-\mu N_3} \, .
%= \sum_{N_3} e^{-\mu 
%N_3}\sum_{\{H_{N_3}\in [\pi_1]\}} {\bf \Upsilon}(H_{N_3},\alpha) \, .
\label{Z2} 
\ee
\nnl This partition function was used in \cite{Bo} to prove
that  the number of simplicial 3-manifolds having a 
fixed homology type grows exponentially with the number of 
tetrahedra, as mentioned above. 
{\it A priori} it is not clear if ${\bf Z}^{(2)}(\alpha,\mu)$ defines 
a theory in the same universality class as the one defined by 
${\bf Z}^{(1)}(\alpha,\mu)$. 
However, ${\bf Z}^{(2)}(\alpha,\mu)$ is the partition 
function of a set of Gaussian fields coupled to geometry in a well-defined 
way, and it is from this point of view equally interesting as 
the system defined by ${\bf Z}^{(1)}(\alpha,\mu)$. 
In the next section we consider a reduced version 
of (\ref{Z2}).

%Naively, one would expect
%the model to be in the same universality class as the model given 
%by (\ref{Z1}). In the next section we consider a reduced version 
%of (\ref{Z2}).

%--------------------------------------------------------------

%\nnl where $\sum_{\{H_{N_3}\}}$ goes over all balanced presentations from a 
%given homotopy class: $\{H_{N_3}=\mv{X_{N_3+1}|R_{N_3+1}}\in 
%[\pi_1]\}$.  
%\be 
%{\bf \Upsilon}(H_{N_3},\alpha) = \sum_{N_0}  
%\sum_{\{P_{N_3+1,N_3+N_0}\cong H_{N_3}\}} e^{\alpha N_0} =  
%\sum_{N_0} e^{\alpha N_0} \Upsilon_{N_0}(H_{N_3})  
%\ee  

%\nnl where $\sum_{\{P_{N_3+1,N_3+N_0}\cong H_{N_3}\}}$ is the sum over all 
%nonreduced presentations $\mv{X_{N_3+1}|R_{N_3+N_0}}$ which can be deduced 
%from a given balanced presentation $H_{N_3}=\mv{X_{N_3+1}|R_{N_3+1}}$ 
%following a pattern of a spanning tree in some simplicial complex. The last 
%equality is a definition of $\Upsilon_{N_0}(H_{N_3})$. 

%%%%%%%%%%%%%%%%% SECTION 2 %%%%%%%%%%%%%%%%%%%%%%% 

\newsection{Reduced model and connection with the 2-dimensional loop gas model} 
In 3 dimensions there are several natural 
choices of classes of complexes one can use in the definition
of the partition function, as already discussed, and  
the question of universality classes is open. 
A simple choice is the set of 3-spheres with imbedded so-called 
2-collapsible spines. A spine is the 2-skeleton, $K_2$, of a complex
$C$ in a cell-decomposition with one 3-cell. In the simplicial context
 this 3-cell is a ball $B_3$ with a triangulated boundary.
Thus we are lead to study the following partition function:
\be  
{\bf Z}^{(2)}_{\rm red}(\alpha,\mu) 
\ = \  \sum_{C \sim S^3}
\sum_{K_2 \subset C}   
e^{\alpha N_0 - \mu N_3} \sum_{T \subset K_2} \, ,
\label{Z2red}
\ee  
where $\sum_{C \sim S^3}$ is the sum over all simplicial 3-spheres,
where $\sum_{K_2 \subset C}$ is the sum over 2-collapsible spines, $K_2$,
embedded in $C$, and where $\sum_{T \subset K_2}$ is the sum over all 
spanning trees in $K_2$. The terms in (\ref{Z2red}) is a subset of the terms 
appearing in (\ref{Z2}) when rewriting the determinant using a trick 
similar to the one used in   (\ref{Z1}) (see \cite{Bo} for details).
The spine $K_2$ in (\ref{Z2red}) is obtained from the
3-sphere $C$ a  pair-wise identification (gluing) of all the 
boundary triangles.  
If $K_2$ is 2-collapsible, the pair-wise identification can be made 
``local'' such that one can use 
recursively the gluing operation consisting of 
the identification of 2 triangles 
sharing a common link on the boundary, or, in other words, a folding along the 
link. In the dual language, one starts with an arbitrary spherical 3-valent 
fat graph and applies the move which can be represented as the flip of a 
link with the subsequent elimination of it:  
\setlength{\unitlength}{1mm} 
\be 
\bp(80,20)(-40,-10) \thicklines \put(-35,0){\line(-1,1){8}} 
\put(-35,0){\line(-1,-1){8}} \put(-35,0){\line(1,0){13}} 
\put(-22,0){\line(1,-1){8}} \put(-22,0){\line(1,1){8}} 
\put(-11,-5){\makebox(10,10){$\Longrightarrow$}} \put(5,5){\line(-1,1){5}} 
\put(5,5){\line(1,1){5}} \put(5,-5){\line(-1,-1){5}} 
\put(5,-5){\line(1,-1){5}} \multiput(5,-5)(0,1){10}{\line(0,1){0.5}} 
\put(15,-5){\makebox(10,10){$\Longrightarrow$}} 
\put(32,10){\oval(10,10)[b]} \put(32,-10){\oval(10,10)[t]} \ep 
\label{move1} 
\ee 
\nnl Both of these moves are well known and have been used in Monte-Carlo 
simulations of triangulated surfaces \cite{KKM}. Having glued all 
triangles, one finishes with a collection of self-avoiding closed loops, 
the number of which equals $N_0-1$. 

It is convenient, instead of erasing a link after a flip, to decorate it 
with a dashed propagator and keep  track of it  in the process of subsequent 
foldings. To represent such configurations, we need to introduce the 
infinite set of vertices 
\be 
\bp(130,20)(5,-10) 
\thicklines 
\multiput(10,0)(30,0){4}{\line(1,0){10}} 
\multiput(15,0)(0,1){10}{\line(0,1){0.5}} 
\multiput(45,0)(0,1){10}{\line(0,1){0.5}} 
\multiput(75,0)(0,1){10}{\line(0,1){0.5}} 
\multiput(105,0)(0,1){10}{\line(0,1){0.5}} 
\multiput(45,0)(0,-1){10}{\line(0,1){0.5}} 
\multiput(75,0)(-0.5,-1){10}{\line(0,-1){0.5}} 
\multiput(75,0)(0.5,-1){10}{\line(0,-1){0.5}} 
\multiput(105,0)(0,-1){10}{\line(0,1){0.5}} 
\multiput(105,0)(-0.5,-1){10}{\line(0,-1){0.5}} 
\multiput(105,0)(0.5,-1){10}{\line(0,-1){0.5}} 
\put(125,-5){\makebox(10,10){and so on}} 
\ep 
\label{vert} 
\ee 
\noindent 
which can be generated by successive flips. For example,\\ 
\noindent 
\bp(25,20)(-10,-10) 
\thicklines 
\put(-5,0){\line(1,0){10}} 
\put(-5,0){\line(-1,-1){5}} 
\put(-5,0){\line(-1,1){5}} 
\put(5,0){\line(1,-1){5}} 
\put(5,0){\line(1,1){5}} 
\multiput(0,0)(0,-1){8}{\line(0,-1){0.5}} 
\ep 
\raisebox{9mm}{gives} 
\bp(18,20)(-10,-10) 
\thicklines 
\put(-5,7){\line(1,0){10}} 
\put(-5,0){\line(1,0){10}} 
\multiput(0,0)(0,1){8}{\line(0,1){0.5}} 
\multiput(0,0)(0,-1){8}{\line(0,-1){0.5}} 
\ep 
\raisebox{9mm}{\Large ;\hspace{5mm}} 
\bp(25,20)(-10,-10) 
\thicklines 
\put(-5,0){\line(1,0){10}} 
\put(-5,0){\line(-1,-1){5}} 
\put(-5,0){\line(-1,1){5}} 
\put(5,0){\line(1,-1){5}} 
\put(5,0){\line(1,1){5}} 
\multiput(-2,0)(0,-1){8}{\line(0,-1){0.5}} 
\multiput(2,0)(0,-1){8}{\line(0,-1){0.5}} 
\ep 
\raisebox{9mm}{gives} 
\bp(20,20)(-10,-10) 
\thicklines 
\put(-5,7){\line(1,0){10}} 
\put(-5,0){\line(1,0){10}} 
\multiput(0,0)(0,1){8}{\line(0,1){0.5}} 
\multiput(0,0)(0.5,-1){8}{\line(0,-1){0.5}} 
\multiput(0,0)(-0.5,-1){8}{\line(0,-1){0.5}} 
\ep 
\raisebox{9mm}{\Large ;\hspace{5mm}} 
%\raisebox{9mm}{and so on} 

\noindent 
\bp(22,20)(-10,-10) 
\thicklines 
\put(-5,0){\line(1,0){10}} 
\put(-5,0){\line(-1,-1){5}} 
\put(-5,0){\line(-1,1){5}} 
\put(5,0){\line(1,-1){5}} 
\put(5,0){\line(1,1){5}} 
\multiput(-2,0)(0,-1){8}{\line(0,-1){0.5}} 
\multiput(2,0)(0,1){8}{\line(0,-1){0.5}} 
\ep 
\raisebox{9mm}{or}  
\bp(22,20)(-10,-10) 
\thicklines 
\put(-5,0){\line(1,0){10}} 
\put(-5,0){\line(-1,-1){5}} 
\put(-5,0){\line(-1,1){5}} 
\put(5,0){\line(1,-1){5}} 
\put(5,0){\line(1,1){5}} 
\multiput(-2,0)(0,1){8}{\line(0,-1){0.5}} 
\multiput(2,0)(0,-1){8}{\line(0,-1){0.5}} 
\ep 
\raisebox{9mm}{or}  
\bp(22,20)(-10,-10) 
\thicklines 
\put(-5,0){\line(1,0){10}} 
\put(-5,0){\line(-1,-1){5}} 
\put(-5,0){\line(-1,1){5}} 
\put(5,0){\line(1,-1){5}} 
\put(5,0){\line(1,1){5}} 
\multiput(0,0)(0,1){8}{\line(0,-1){0.5}} 
\multiput(0,0)(0,-1){8}{\line(0,-1){0.5}} 
\ep 
\raisebox{9mm}{give} 
\bp(20,20)(-10,-10) 
\thicklines 
\put(-5,3){\line(1,0){10}} 
\put(-5,-3){\line(1,0){10}} 
\multiput(0,-10)(0,1){20}{\line(0,1){0.5}} 
\ep 
\raisebox{9mm}{\Large ;\hspace{5mm}} 

\noindent 
\bp(22,20)(-10,-10) 
\thicklines 
\put(-5,0){\line(1,0){10}} 
\put(-5,0){\line(-1,-1){5}} 
\put(-5,0){\line(-1,1){5}} 
\put(5,0){\line(1,-1){5}} 
\put(5,0){\line(1,1){5}} 
\multiput(-3,0)(0,-1){8}{\line(0,-1){0.5}} 
\multiput(0,0)(0,-1){8}{\line(0,-1){0.5}} 
\multiput(3,0)(0,1){8}{\line(0,-1){0.5}} 
\ep 
\raisebox{9mm}{or}  
\bp(22,20)(-10,-10) 
\thicklines 
\put(-5,0){\line(1,0){10}} 
\put(-5,0){\line(-1,-1){5}} 
\put(-5,0){\line(-1,1){5}} 
\put(5,0){\line(1,-1){5}} 
\put(5,0){\line(1,1){5}} 
\multiput(-3,0)(0,-1){8}{\line(0,-1){0.5}} 
\multiput(0,0)(0,1){8}{\line(0,-1){0.5}} 
\multiput(3,0)(0,-1){8}{\line(0,-1){0.5}} 
\ep 
\raisebox{9mm}{or}  
\bp(22,20)(-10,-10) 
\thicklines 
\put(-5,0){\line(1,0){10}} 
\put(-5,0){\line(-1,-1){5}} 
\put(-5,0){\line(-1,1){5}} 
\put(5,0){\line(1,-1){5}} 
\put(5,0){\line(1,1){5}} 
\multiput(-3,0)(0,1){8}{\line(0,-1){0.5}} 
\multiput(0,0)(0,-1){8}{\line(0,-1){0.5}} 
\multiput(3,0)(0,-1){8}{\line(0,-1){0.5}} 
\ep 
\raisebox{9mm}{or}  
\bp(22,20)(-10,-10) 
\thicklines 
\put(-5,0){\line(1,0){10}} 
\put(-5,0){\line(-1,-1){5}} 
\put(-5,0){\line(-1,1){5}} 
\put(5,0){\line(1,-1){5}} 
\put(5,0){\line(1,1){5}} 
\multiput(0,0)(-0.5,-1){8}{\line(0,-1){0.5}} 
\multiput(0,0)(0.5,-1){8}{\line(0,-1){0.5}} 
\multiput(0,0)(0,1){8}{\line(0,-1){0.5}} 
\ep 
\raisebox{9mm}{give}  
\bp(20,20)(-10,-8) 
\thicklines 
\put(-5,6){\line(1,0){10}} 
\put(-5,0){\line(1,0){10}} 
\multiput(0,0)(0,1){12}{\line(0,1){0.5}} 
\multiput(0,0)(0.5,-1){8}{\line(0,-1){0.5}} 
\multiput(0,0)(-0.5,-1){8}{\line(0,-1){0.5}} 
\ep 
\raisebox{9mm}{\Large ,\hspace{5mm}}  

\noindent 
and so on. 

In the end, we obtain planar diagrams with two types of propagators: solid 
ones produce closed loops while dashed form arbitrary clusters 
of dotted lines. The total 
number of vertices is equal to $2(N_3+1)$. We weight every loop with 
the numerical factor $e^{\alpha}$. 

Thus, the model can be represented as a non-local gas of closed loops with the 
partition function 
\be 
{\bf Z}^{(2)}_{\rm red}(\alpha,\mu)
\ = \ \sum_{\{{\cal D}\}} w_{\cal D} \, e^{\alpha N_0-\mu N_3} \, ,
\label{nlocLG} 
\ee 
\nnl where $\sum_{\{{\cal D}\}}$ is the sum over all the diagrams. The 
weight $w_{\cal D}$ is equal to the number of inequivalent starting 
configurations giving the same planar diagram $\cal D$ (can be 0). The 
number of closed loops is simply equal to $N_0-1$ and $\alpha$ can be 
identified with the inverse Newton constant.

We see that the weights $w_{\cal D}$ in \eq{nlocLG} in principle 
are recursively calculable. 
However, from a practical point of view 
it is difficult to  take into account  the entropy 
of the configurations, and the model (\ref{nlocLG}) seems to be analyticlly 
unsolvable. Therefore, let us consider the localized version of the model, 
namely, the dense phase of the self-avoiding loop gas matrix model (in the 
following simply denoted the loop gas (LG) model): 
\begin{eqnarray}
Z_{\rm LG} &=& \int d^{\scs N^2}Y \prod_{\nu=1}^n d^{\scs N^2}X_{\nu}  
\exp\Big[-\frac{N}2\tr Y^2-\frac{N}2\sum_{\nu=1}^{n}\tr X^2_{\nu}  
+\frac{N}2\sum_{\nu=1}^{n}\tr YX^2_{\nu}\Big] 
\nonumber\\ %\]\be 
&=& \int d^{\scs N^2}Y 
\exp\Big[-\frac{N}2\tr Y^2 -\frac{n}2\tr^2\log\Big(1\otimes 1 
- \mu Y\otimes 1 -  1 \otimes \mu Y \Big) \Big] \, .
\label{LG}
\end{eqnarray}
Here $Y$ and $X_{\nu}$ are hermitian $N\times N$ matrices. 
We have attached the lower index to the gaussian $X_{\nu}$ variable  
to weight every closed loop with the factor $n\equiv e^{\alpha}$.  
In the $N\to\infty$ limit, this model generates planar diagram having only 
the simplest 3-valent vertices from the whole series (\ref{vert}). For such 
diagrams, all the weights in \eq{nlocLG} are trivial: $w_{\cal D}=1$. 

This truncated local model might have some qualitative features 
in common with 3-dimensional simplicial gravity. 
It seems to be a reasonable assumption, 
because the corresponding 
universality class is very large, (see \rf{O(n)}),  
and includes all local perturbations of 
the model (\ref{LG}). On the other hand, multiple overlapping sequences of 
the flips can be regarded as a kind of renormalization group 
procedure, which could draw any given system to a stable fixed point in 
the space of all planar-graph models. It looks a bit more natural in 
dual terms. One takes an arbitrary spherical ball with a huge triangulated 
boundary and uses subsequently and randomly the folding operation (dual to 
the flip). Soon the boundary triangulation will be randomized. If this 
randomization has appropriate robust statistical properties, we would 
expect to find many models falling in the same universality 
class. 

Then any 2 models from the 
same universality class should be identical in their continuum limits. 

The critical behavior of the loop gas matrix model is well known.  For 
$n<2$, it describes 2D gravity interacting with $c<1$ conformal matter 
\be 
c \ = \ 1 - 6\frac{(1-g_0)^2}{g_0}; \hspace{2pc} 
n \ = \ -2 \cos \pi g_0 \, ,
\ee
while for $n>2$ the  
corresponding matter is non-critical and the model trivializes
except in special situations \cite{charlotte}. 
In this phase, the number of closed loops is proportional to the volume  
and the mean length of each remains finite. 

In the vicinity of a critical point $\mu_c$, the partition function  
behaves as 
\be  
Z_{\rm LG} \ \approx \ (\mu_c-\mu)^{2-\gamma_{\rm str}} \, .
\label{defgam} 
\ee  
This formula 
defines the famous string susceptibility exponent $\gamma_{\rm str}$. 

Two main quantities of interest are 
\be 
\mv{N_3}_{LG} 
\ = \ \frac{\d\ }{\d\mu} \log Z_{\rm LG} 
\ \approx \ -(2-\gamma_{\rm str}) \frac{\mu_c}{\mu_c-\mu} 
\ee 
and 
\begin{eqnarray}%\[ 
\mv{N_0}_{LG} 
\ = \ \frac{\d\ }{\d\alpha} \log Z_{\scs LG} 
&\approx& (2-\gamma_{\rm str}) \frac{\mu'_c(n)}{\mu_c(n)-\mu} 
  \, - \, \frac{\d\gamma_{\rm str}} {\d\alpha}\log|\mu_c-\mu| 
\nonumber\\ %\]\be 
&\approx& - \frac{\mu'_c}{\mu_c}\mv{N_3}_{LG} 
    \, + \, \frac{\d\gamma_{\rm str}} {\d\alpha}\log\mv{N_3}_{LG} 
\end{eqnarray} %\ee
where the coefficients can be calculated explicitly using the Gaudin and  
Kostov's exact solution \cite{O(n)} 
\be 
\frac{\d\ }{\d n}\log\mu_c \ = \ - \frac12\frac1{n+2}; 
\hspace{2pc} 
\frac{\d\gamma_{\rm str}}{\d n} \ = \ \frac1{2\pi g_0^2 \sin\pi(1-g_0)} \, .
\ee 
If $2-n\ll 1$, then 
\be 
\mv{N_0}_{LG} \ \approx \ \frac14\mv{N_3}_{LG}  
\, + \, \frac1{\pi\sqrt{2-n}}\log\mv{N_3}_{LG} \, .
\ee
{\noindent}
If $n>2$, then $\gamma_{\rm str}$ is independent of $n$ and  
$\mv{N_0}_{LG}$ is proportional to $\mv{N_3}_{LG}$, except 
in special situations \cite{charlotte}.

\newsection{Numerical experiments} 

Unfortunately, the nonlocal model (\ref{nlocLG}) looks analytically 
unsolvable. However, it can be successfully simulated numerically by the 
recursive sampling-Monte-Carlo technique. It is natural to use a 
recursive sampling to perform 
the summation over the tree-dual balls in a close analogy with 
Ref.~\cite{KKSW}. Then, a Metropolis algorithm can be used to produce a 
random walk in a space of all simplicial spherical balls. Closed manifolds 
appear when the boundary of a ball is shrunk, again by a trial and error
Monte Carlo procedure. In the following we describe the algorithm in 
some detail.

\subsection{The algorithm for generating three-spheres}

As mentioned, a  recursive algorithm has been used with 
great success for a $c=-2$ model 
coupled to two-dimensional quantum gravity. 
Since our model resembles the $c=-2$ model quite a lot, 
it is natural to try to use the same technique in three dimensions. 
One of the main reasons for the success of the algorithm 
in the two-dimensional case was that 
the branching probability of tree-graphs and the counting of 
rainbow diagrams could be calculated analytically. 
In the present three-dimensional model 
the branching probability can be calculated analytically, 
as shown below. 
However, while there is a three-dimensional analogy of 
the rainbow diagram encountered in two dimensions, 
they are not two-dimensional extended objects and 
we do not know how to count them.

Let us calculate the branching probability of 3-valent tree-diagrams.
The Schwinger-Dyson equation for 3-valent graphs corresponding to 
Fig.~1 is 
\be\label{qbic}
T \ = \ xT^3 +1,
\ee
where $T(x)$ is the generating function for the number of 
tree diagrams, i.e., 
\[
T(x) \ = \ \sum_n t_nx^n,
\]
where $t_n$ is the number of 3-valent tree-diagrams with $n$ vertices.

%%%%%%%%%%%%%%%%%%%%%%%%%%%%%%%

\begin{figure}
\centerline{\hbox{\psfig{figure=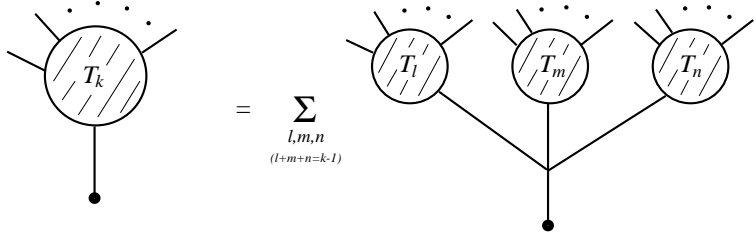,width=10cm,angle=270}}}
%\centerline{\hbox{\psfig{figure=tree.ps,height=8cm,width=10cm,angle=90}}}
\caption[fig_tree]{Graphical presentation of Schwinger-Dyson equation 
for 3-valent tree-diagrams.} 
\label{fig_tree}
\end{figure}

%%%%%%%%%%%%%%%%%%%%%%%%%%%%%%%

Since the Schwinger-Dyson equation is a cubic equation 
with respect to $T$, it has three solutions. We have to select 
the unique solution which satisfies the physically imposed 
boundary condition $t_0 =1$. This results in the following solution:
\be\label{sol}
T \ = \ \frac{3}{z} \, \sinh \Bigl( \frac{1}{3} 
        \log ( z+ \sqrt{1+z^2}) \Bigr),
~~~~
z = i \sqrt{\frac{27x}{4}},
\ee
which has the following power series expansion:
\be\label{series}
T(x) \ = \ \sum_{n=0}^\infty t_nx^n 
\ = \ \sum_{n=0}^{\infty} \frac{(3n)!}{n!(2n+1)!}\, x^n.
\ee

We can now define the probability of the 3-valent tree-diagrams with 
$k$ vertices branching into tree-diagrams with $l$, $m$ and $n$ vertices
\be
w(l,m,n) \ = \ \frac{t_lt_mt_n}{t_k}~~~~~~~(l+m+n=k-1),
\ee 
which then naturally satisfies
\be
\sum w(l,m,n) \ = \ 1,
\ee
where the summation is over $l,m,n \geq 0$ with the constraint 
$l+m+n=k-1$. By rearranging the summation we obtain the following relation
\be
w(l,m,n) \ = \ w_1(l)\, w_2(m,n),
\ee
where 
\be
w_1(l) = \hspace{-0.75cm}\sum_{\mbox{\scriptsize{$
\matrix{m',n' \geq 0 \cr m'+n'=k-l-1}$}}}\hspace{-0.75cm} 
w(l,m',n'),
~~~~w_2(m,n) = \frac{w(l,m,n)}{w_1(l)},
\ee
where $m+n= k-l-1$. Since we have the relations
\be
\sum_{l=0}^{k-1} w_1(l) \ = \ 1,~~~~~
\sum_{\mbox{\scriptsize{$\matrix{m,n \geq 0 \cr m+n=k-l-1}$}}}
\hspace{-0.75cm} w_2(m,n) \ = \ 1,
\ee
we can divide the branching process into two parts using the relation 
\be
\sum_{\mbox{\scriptsize{
$\matrix{m,n \geq 0 \cr m+n=k-l-1}$}}}\hspace{-0.75cm} w(l,m,n) \ = \ 
\frac{3(k-l)-2}{k-l}\; 
\frac{t_l t_{k-l-1}}{t_k}~~~~~(0 \leq l \leq k-1),
\ee
and write:
\be
w_1(l) \ = \ \frac{3(k-l)-2}{k-l} \; \frac{t_lt_{k-l-1}}{t_k},~~~~
w_2(m,n) \ = \ \frac{k-l}{3(k-l)-2}\; \frac{t_mt_n}{t_{m+n}}.
\ee
Using these analytic formulas we can set up a recursive sampling 
algorithm for constructing 3-valent tree-diagrams in three dimensions.

In order to construct a three-dimensional simplicial manifold
with the topology of $S^3$ from the 3-valent diagrams, we need 
to identify the ``end branches'' of the tree-diagrams 
in the dual diagram sense, or equivalently, successive neighboring
triangles of the corresponding original manifold. In contrast to the 
two-dimensional $c=-2$ model, where an analytic formula for the 
number of rainbow diagrams is given, it is presently unknown how 
to obtain a closed expression in the three-dimensional case 
since the trees are now genuine three-dimensional objects.
The initial three-dimensional tree diagram corresponds to the tree shape 
volume object which has a ball topology with a triangulated $S^2$ sphere 
as a boundary. It should be noted that in this model all the links of the 
corresponding three-dimensional simplicial manifold are on the boundary 
in the initial tree configuration. 
We first randomly pick up a link on the boundary and identify the neighboring 
triangles which share the link as a common one.
This link is now hidden inside from the boundary. 
We then pick up another link which is on the boundary surface and 
repeat the gluing procedure of the triangles.
This procedure should 
be continued successively until all the triangles on the boundary surface 
are identified (and the boundary thereby has vanished).
Since we don't know the analytic expression of the weight factor of the 
three-dimensional \lq\lq rainbow diagrams", we need to introduce the ungluing 
process and use a Monte Carlo simulation to obtain a typically triangulated 
three-dimensional $S^3$ sphere in our model.
Since the gluing 
process is carried out by a Monte Carlo algorithm, we operate 
both with the gluing process and its reciprocal in such a way 
that the usual detailed balance is fulfilled.
In this procedure the order of gluing and ungluing processes, 
which are reflected in the internal configuration of the simplicial 
manifold, are important and thus should be memorized in the simulation.
In this Monte Carlo procedure there is unavoidably introduced 
an iteration uncertainty in the counting procedure.

\subsection{Observables}

After constructing typical $S^3$ manifolds one can use these
to calculate expectation values of observables.
A basic observable which probes the phase diagram of the model is the 
``string'' susceptibility exponent $\gamma$ defined analogously to 
Eq.~(\ref{defgam}). The most convenient way to extract $\gamma$
is by means of the so-called baby universe technique \cite{A2},
which contrary to the original definition (\ref{defgam}) allows
us to extract $\gamma$ from an ensemble of manifolds
with fixed volume $N_3$. 
One counts the configurations of three-manifolds which are 
separated into two parts via a common triangle, and then measures
the volumes $M_3$ and $N_3-M_3$ of the separated parts. The 
string susceptibility is then derived from the following relation:
\be\label{baby}
W(M_3) \ = \ 
\frac{M_3 Z(M_3)\; (N_3-M_3)Z(N_3-M_3)}{Z(N_3)} \ \sim \
N_3\left[\frac{M_3}{N_3}  
\Big(1- \frac{M_3}{N_3}\Big)\right]^{\gamma-2},
\ee
where the partition function $Z(N_3)$ (in accordance with (\ref{defgam})) 
is assumed to be parametrized as 
$Z(N_3) \sim N_3^{\gamma-3}\;e^{-\mu_c N_3}$.  
In particular we obtain the following approximate scaling function: 
\be 
F(x) \ = \ N_3 \log\frac{W(M_3+1)}{W(M_3)}  \ \sim \ 
(\gamma - 2)\left( \frac{1}{x} - \frac{1}{1-x}\right),  
\label{fitcurv} 
\ee
where the scaling parameter is $x = M_3/N_3$.

%%%%%%%%%%%%%%%%%%%%%%%%%%%%%%%

\begin{figure}
\centerline{\hbox{\psfig{figure=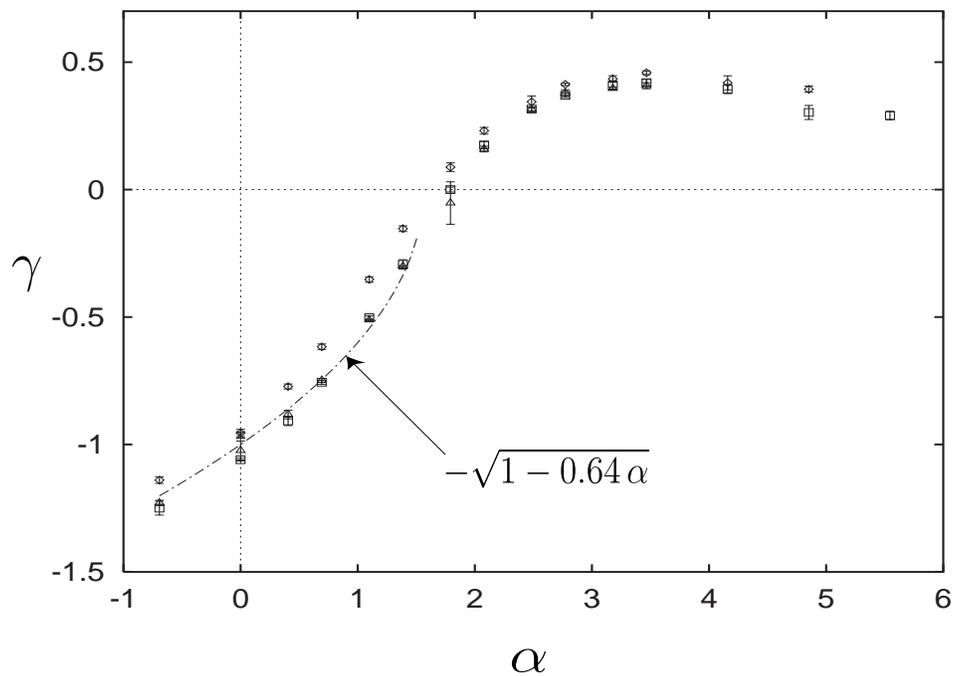,height=9.0cm,width=13.5cm,angle=0}}}
\centerline{\hbox{\psfig{figure=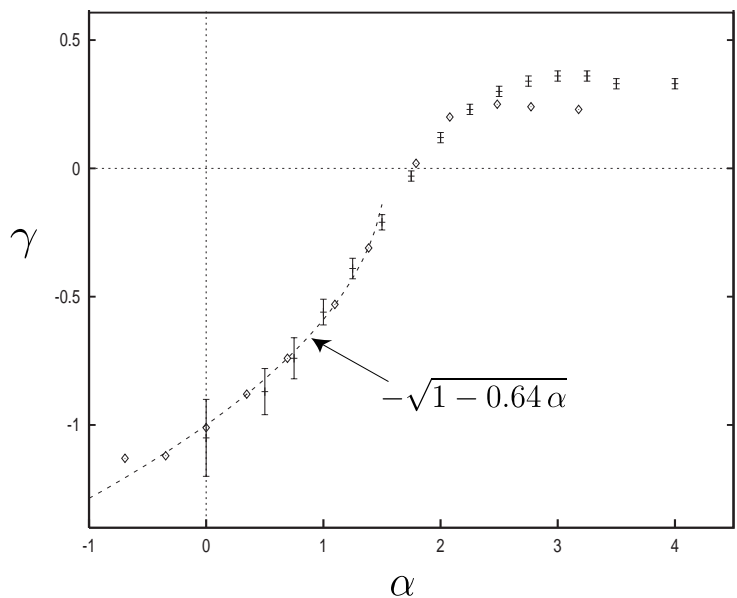,height=9.1cm,width=10.1cm,angle=0}}\hspace*{1.8cm}}
\caption[fig_tree]{``String" susceptibility $\gamma$ {\rm versus} $\alpha$, 
the inverse Newton constant, for $N_3 = 50$($\diamond$), 
$100$($-$), $200$($\Box$), $300$($\times$) %($\alpha=0$) 
above and $N_3 = 300$($\diamond$), $500$($-$) below.}
\label{fig_alldata}
\end{figure}
%%%%%%%%%%%%%%%%%%%%%%%%%%%%%%%

%%%%%%%%%%%%%%%%%%%%%%%%%%%%%%%

%\begin{figure}
%\centerline{\hbox{\psfig{figure=gamma_alpha.eps,width=10cm,angle=0}}}
%\caption[fig_tree]{``String" susceptibility $\gamma$ {\rm versus} $\alpha$, 
%for $N_3$ = 300,500.} 
%\label{fig_alldata}
%\end{figure}

%%%%%%%%%%%%%%%%%%%%%%%%%%%%%%%

\begin{figure}
\centerline{\hbox{\psfig{figure=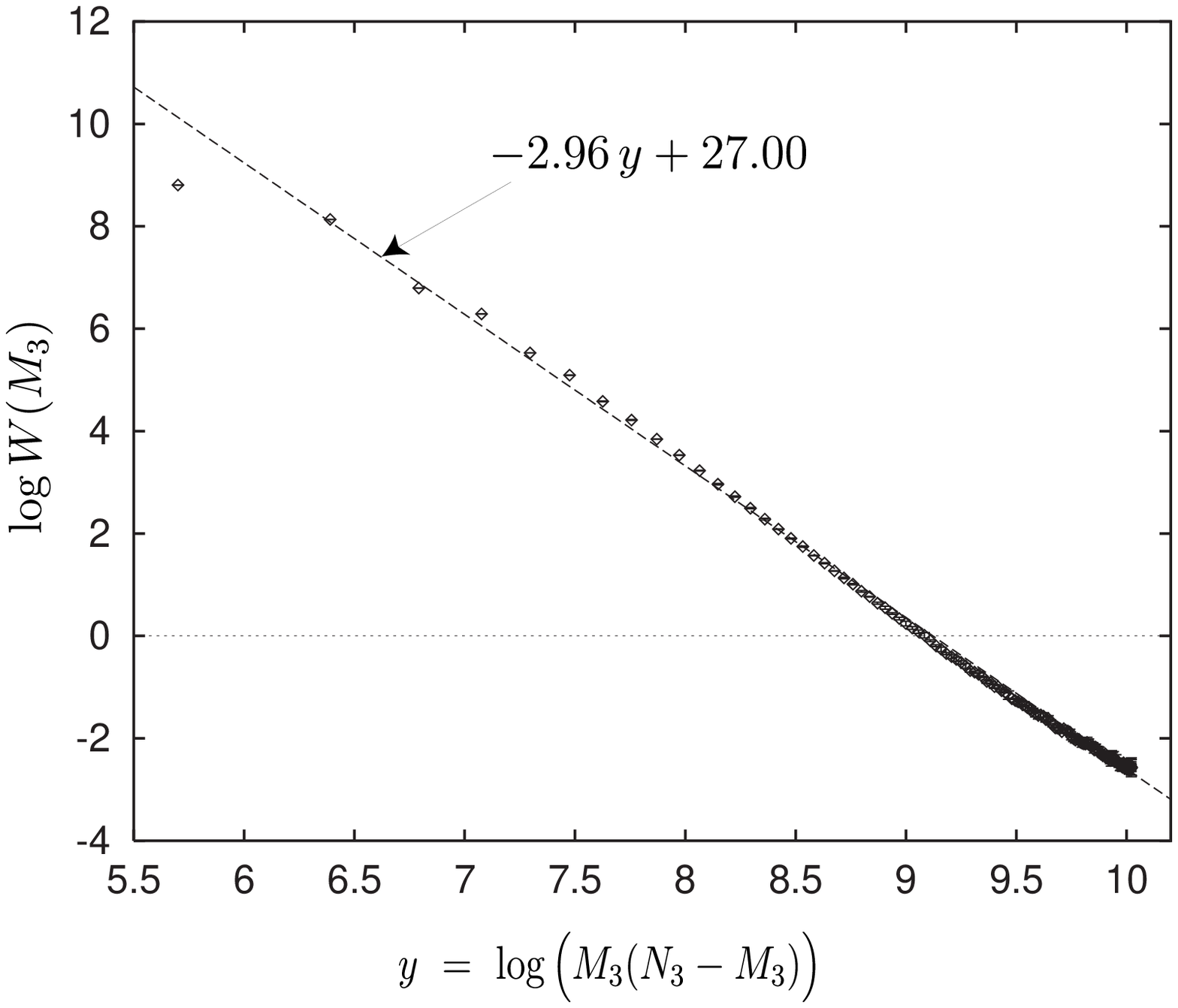,width=10cm,angle=0}}}
\caption[fig_gammaslope]{$\log W(M_3)$ {\rm versus} 
$\log\big(M_3(N_3-M_3)\big)$ at $\alpha = 0$ for $N_3 = 300$.} 
\label{fig_gammaslope}
\end{figure}

A number of measured values of $\gamma$ as a function 
of $\alpha$, the inverse Newton constant, is shown in Fig.~\ref{fig_alldata}. 
Statistical error-bars are shown in the Figure. 
The two sub-figures in Fig.~\ref{fig_alldata} represent 
independent measurements 
for the different value of $N_3$ and they agree perfectly within error-bars.
Systematic errors (\eg due to finite 
size effects) seem to be of the same order except in a vicinity of the 
transition point.  
The data are collections of three simulations which took place approximately 
4-6 months of CPU time for each calculation on a Pentium 200 MHz personal 
computer and HP Convex and Origin 2000 parallel computers. 
We have collected data  for 
$N_3=50$, $100$, $200$ at $\log(1/2) \le \alpha \le \log(256)$, 
and $N_3= 300$ at $\alpha=0$ and have shown the extracted $\gamma$ 
in the upper sub-figure of Fig.\ \ref{fig_alldata}. Other data 
for $N_3=300$, $500$ lead the $\gamma$-values shown in 
the lower sub-figure.
Although the number of tetrahedra are small, 
they seem to give reliable measurements of $\gamma$ 
in the interval $-0.7<\alpha<5.5$, except for $N_3=50$. 
For the value of $N_3$ there is a systematic deviation 
from the other values, 
probably caused by the small number of $N_3$. 
A typical distribution of $\log W(M_3)$ for $\alpha = 0$ and $N_3=300$ 
is shown in Fig.~\ref{fig_gammaslope} 
where the slope of the curve determines $\gamma$.

%along with a fitting curve of the 
%form (\ref{fitcurv}). 

%%%%%%%%%%%%%%%%%%%%%%%%%%%%%%%

%%%%%%%%%%%%%%%%%%%%%%%%%%%%%%%

%%%%%%%%%%%%%%%%%%%%%%%%%%%%%%%

\begin{figure}
\centerline{\hbox{\psfig{figure=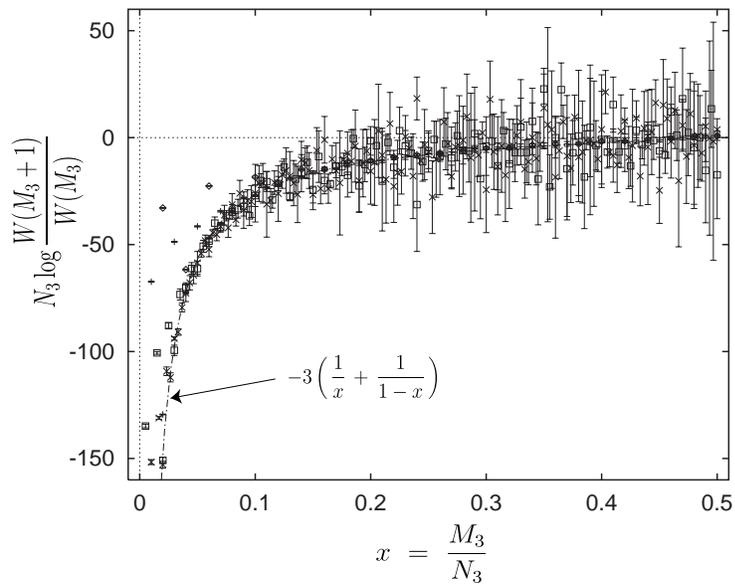,width=10cm,angle=0}}}
\caption[fig_scaling]{Scaling function $F(x)$ {\rm versus} $x=M_3/N_3$ at 
$\alpha=0$ for $N_3 = 50$, $100$, $200$, $300$.} 
\label{fig_scaling}
\end{figure}

%%%%%%%%%%%%%%%%%%%%%%%%%%%%%%%

\begin{figure}
\centerline{\hbox{\psfig{figure=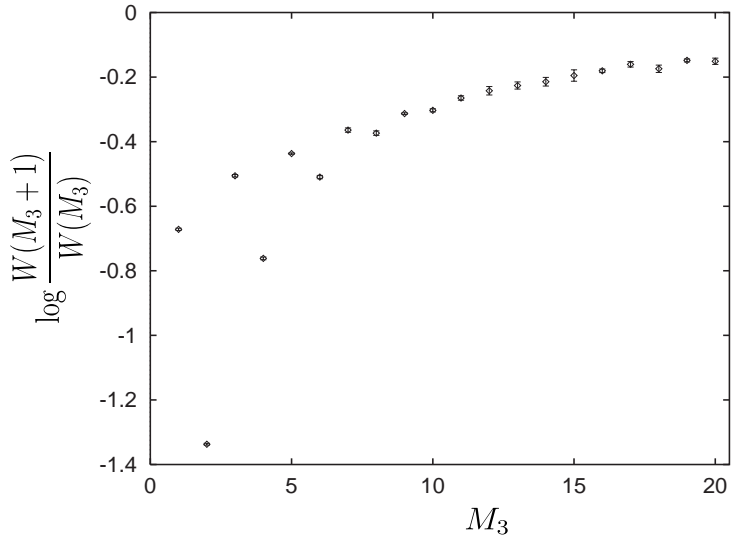,width=10cm,angle=0}}}
\caption[fig_oscilation]{$\hbox{log} \frac{W(M_3+1)}{W(M_3)}$ 
{\rm versus} $M_3$ at $\alpha$ = 0 for $N_3$ =300.} 
\label{fig_oscilation}
\end{figure}

The scaling function $F(x)$ with the predicted curve (\ref{fitcurv}) 
is shown in Fig.~\ref{fig_scaling} 
for $\alpha = 0$ with $N_3=50$, $100$, $200$, $300$.
We can see that all the data scale nicely within the error-bars
except for $M_3 <\hspace*{-4 mm}\raisebox{-2 mm}{$\sim$}~ M_3^{\rm min} = 5$.
As the formula (\ref{fitcurv}) is valid only asymptotically, 
we can use fitting only for  values of $M_3$ bigger than some 
cutoff $M_3^{\rm min}$. 
The 
independence of results from this cutoff can serve as an indicator of 
the reliability. 
A corresponding typical measurements 
are shown in Fig.~\ref{fig_oscilation}. For small 
values of 
$M_3 <\hspace*{-5 mm}\raisebox{-2 mm}{$\sim$}~ M_3^{\rm min} = 5$, the 
measured ``virtual'' 
values of $\gamma$ 
oscillate and for larger values they gradually converge 
to a stable limit.  

%%%%%%%%%%%%%%%%%%%%%%%%%%%%%%%

%%%%%%%%%%%%%%%%%%%%%%%%%%%%%%%

In the measurements of $\gamma$ we have thus collected the data 
satisfying 
$M_3 >\hspace*{-4 mm}\raisebox{-2 mm}{$\sim$} ~M_3^{\rm min} + 1 = 6$.
We have measured $\gamma$ with good statistics for 
$\alpha = 0$ where we collected $10^5$ configurations while for the other 
values we collected about $2\times 10^4$ configurations. 
The measured value $\gamma$ is so close to $-1$ that it is 
tempting to conjecture that this is the exact value for $\alpha =0$.
One could hope that it was possible to prove this 
result analytically, since it corresponds to the 
strong coupling limit $G = 1/\alpha = \infty$.

%Another plausible explanation of 
%the drift mentioned above is that, for smaller values of $\alpha$, $Z(M_3)$ 
%starts to scale 
%correctly for bigger and bigger values of $M_3$. Then, what is shown in 
%Fig.~\ref{fig_alldata} has to be regarded as virtual values of 
%$\gamma$ as they can be seen at the 
%scale $M_3 \approx10$. 
%To find out which of the two explanations is correct, 
%additional more precise measurements are needed. 
%The role played by finite 
%size effects in the present context is not yet well understood. 

\newsection{Discussion} 

The results of the numerical experiments give some support 
to the idea that there should be a qualitative similarity of the phase 
structures of 2- and 3-dimensional simplicial gravities with  Newton's 
coupling constant playing a role analogous 
to the matter central charge in 2 
dimensions. A similar idea can be found in a slightly 
different context in \cite{mottola}. The experimental 
values of $\gamma$ shown in Fig.~\ref{fig_alldata} support 
the picture of 2 phases with the 
critical coupling $\alpha_{\rm c} \approx 2$. 
For $\alpha < \alpha_{\rm c}$, the simplest square root function 
\be 
\gamma \ = \ - \sqrt{1-0.64\, \alpha\,} 
\label{gamfit} 
\ee 
fits surprisingly well the experimental points. 
We show the above expression by the dotted curve in Fig.~\ref{fig_alldata}.
For $\alpha > \alpha_{\rm c}$, 
$\gamma$ is a constant within error bars about $0.3$. 
Since the systems are quite small, and since there still is some 
drift in the extracted values of $\gamma$ towards larger values 
when $N_3$ is increased (see in particular lower sub-figure in 
Fig.\ \ref{fig_alldata}), we cannot conclude if $\gamma =1/2$ in the 
limit of infinite $N_3$ or if $\gamma = 1/3$, which is the next lowest 
value that $\gamma$ can have in the branched polymer phase \cite{bp}.

From a theoretical point of view, 
it has been recognized by H.\ Kawai and one of authors (Y.W.) 
that in the case of $D > 2$ and around $1/\alpha = G = 0$, 
the branched configurations give much more dominant contribution 
to the path-integral over  metrics 
than the configurations near the flat space-time\cite{KawaiWatabiki}. 
Let us consider the scaling property of 
\begin{equation}\label{X}
X  \ = \  \exp\! \bigg( \frac{1}{G} \int\!\! \sqrt{g} R \bigg) \, .
\end{equation}
Near flat space-time dimensional analysis 
is  reliable and leads to 
%According to the dimensional analysis, 
$\int\!\! \sqrt{g} R \sim V^{1 - \frac{2}{D}}$. 
Then one finds 
\begin{equation}\label{Xflat}
X  \ \sim \  
\exp\! \bigg( \hbox{const.} \frac{V^{1 - \frac{2}{D}}}{G} \bigg) \, ,
\end{equation}
where $V$ is the volume of space-time.
On the other hand, 
for  branched polymer like configurations 
it is easy to show that 
i.e.\ 
$\int\!\! \sqrt{g} R \sim V$. 
Thus one finds in this case that 
\begin{equation}\label{Xbranched}
X  \ \sim \  
\exp\! \bigg( \hbox{const.} \frac{V}{G} \bigg) %\, .
\end{equation}
instead of (\ref{Xflat}). 
Comparing eqs.\ (\ref{Xflat}) and (\ref{Xbranched}) around $G \sim 0$, 
one finds that 
(\ref{Xbranched}) gives a more dominant  contribution 
to the path-integral than the contribution (\ref{Xflat}) from flat space.
This is one reason why one cannot perform the quantization of gravity 
perturbatively around the flat space-time 
(except for the very special cases 
when $1/\alpha_{\rm c} = G_{\rm c} = 0$). 
% like the superstring theories). 
This theoretical observation is consistent with the result of 
this paper. 

If we accept the result that there are two phases in the present 
model of 3-dimensional 
simplicial gravity, then the next natural question to ask 
is: ``How large is the corresponding universality class?'' 
Does the complete model (\ref{Z2}) with a sum over 
all 3-spheres belong to the same class? 
It seems plausible, but it is difficult to judge 
how restricted the subset of simply connected 3-manifolds 
taken into account in the simulated 
statistical ensemble really is. 
It is not easy to construct a simplicial 3-sphere 
which would not be included in it. 
At first thought, Zeeman's Dunce Hat\cite{Zeeman} 
could serve as a mean to construct a counter example. 
Triangles belonging to a boundary of an initial ball 
cannot form such a pseudosurface after local foldings. 
However, those lying inside the ball can. 
Even in the simplest case of 2 tetrahedra, 
a Zeeman's Dunce Hat can easily be obtained.  
Further, the construction given in Ref.~\cite{DJ} is not a 
``counter example'' because it is not a triangulation. 
On the other hand, 
there is an old result of Haken \cite{Haken} who showed that 
any 3-sphere possesses an obviously trivial presentation of 
$\pi_1$ (in the sense that it can be trivialized by a sequence of 
simple substitutions of the length less or equal than 4). 
Unfortunately, this result does not imply that 
any simplicial 3-sphere allows for at least one 2-collapsible spine 
(if it did, the famous Poincar\'{e} hypothesis would 
be simultaneously proven). 
It seems 
an open question whether any simplicial 3-sphere is counted at 
least once in the partition function (\ref{nlocLG}). 
If it is the case, then 
only the matter sector of the original model (\ref{Z2}) 
is distorted by our simplifications leading to the reduced model. 
Even if not all 3-spheres are included in the model, 
it seems reasonable to conjecture that 
the reduced model and the original model belong to 
the same universality class as far as their continuum limits 
are concerned.  

%%%%%%%%%%%%%%%%%%%%%%%%%%%%%%%%%

%\begin{figure}[h]
%\begin{center}
%  \begin{minipage}[c]{.5\textwidth}
%  \epsfxsize=\textwidth \epsfbox{tree.eps}
% \end{minipage}
%\end{center}
%\caption{
%Graphical presentation of geometrical structure of a 
%4-simplex $ABCDE$
%}
% \label{fig:tree}
%\end{figure}

%%%%%%%%%%%%%%%%%%%%%

\bigskip 

\noindent
{\Large\bf Acknowledgements}

\noindent
J.A. acknowledges the support of MaPhySto, Centre of Mathematical 
Physics and Stochastics, which is financed by the Danish National 
Research Foundation. 
This work is supported in part by Japanese Ministry of Education 
under the Grant-in-Aid number 09640330 and the special funding for 
basic research under the project name ``Hierarchical Structure of 
Matter Analyzing system".

\end{document}